\documentclass[prb,aps,twocolumn]{revtex4}
\usepackage{amsmath}
\usepackage{amssymb}
\usepackage {graphicx}

\newcommand{\beq}{\begin{equation}}
\newcommand{\eeq}{\end{equation}}
\newcommand{\bea}{\begin{eqnarray}}
\newcommand{\eea}{\end{eqnarray}}

\newcommand{\D}{\Delta}

\begin{document}

\title{Relation between $2\Delta/T_c$ and nodes in Fe-based superconductors}
\author{Saurabh Maiti, Andrey V. Chubukov}
\affiliation {~Department of Physics, University of Wisconsin,
Madison, Wisconsin 53706, USA}
\date{\today}

\begin{abstract}
We analyze the interplay between the absence or presence of the
nodes in the superconducting gap along electron Fermi surfaces
(FSs) in Fe-pnictides, and  $2\Delta_h/T_c$ along  hole FSs,
measured by ARPES. We solve the set of coupled gap equations for
4-band and 5-band models of Fe-pnictides and relate the presence
of the nodes to $2\Delta_h/T_c$ being below a certain threshold.
Using ARPES data for $2\Delta_h/T_c$, we find that optimally doped
$Ba_{1-x}K_xFe_2As_2$ and $Ba(Fe_{1-x}Co_x)_2As_2$ and likely
nodeless, but iso-valent $BaFe_2(As_{1-x}P_x)_2$ likely has nodes.
\end{abstract}

\maketitle

{\it Introduction}~~~  The properties of Fe-based superconductors
(FeSC) continue to attract interest in the condensed matter
community. Amongst them, the pairing symmetry and the gap
structure are the most debated topics. FeSCs are
multi-orbital/multi-band quasi-2D systems, with 2 cylindrical hole
pockets centered at $(0,0)$, and 2 cylindrical electron pockets at
$(\pi,0)$ and $(0,\pi)$ in the unfolded Brilluoin Zone(BZ), with
one Fe-atom per unit cell. In some systems, there is an additional
cylindrical hole pocket at $(\pi,\pi)$. This FS topology allows
several different gap symmetries and structures: $s^{++}$ and
$s^{\pm}$ s-wave states, $d_{x^2-y^2}$ and $d_{xy}$ states, etc.
\cite{pairing symmetries,d_xy,triplet,Mazin,s++}. Most of
theoretical and experimental studies
\cite{Mazin,graser,Rthomale,dhl_10,zlatko,
andrey,coexistense,spin_resonance} favor a $s\pm$ state in which
the gaps averaged along electron and hole pockets are of opposite
sign. The $s\pm$ gap generally has some $\pm\cos 2 \theta$
variations along the two electron FSs and has accidental nodes
when such a variation is large. Penetration depth and thermal
conductivity data indicate that in some of the FeSCs the gap
probably has no
nodes~\cite{thermal_opt,pen_opt,thermal_pen_nodeless}, while in
the others accidental nodes are likely. The  most compelling
evidence for the nodes is for strongly doped
Ba(Fe$_{1-x}$Co$_x)_2$As$_2$\cite{thermal_opt,Co-over doped},
LaFeAsO$_{1-x}$F$_x$~\cite{lafop} and for iso-valent
BaFe$_2$(As$_{1-x}$P$_x$)$_2$ \cite{P_mSR_ThermalCon}.

The nodes in the gap can be directly probed by angle-resolved
photoemission (ARPES). ARPES measures the gaps in the folded BZ,
where the three hole FSs are all at $(0,0)$ and the two electron
FSs are both at $(\pi,\pi)$. At present, ARPES measurements
distinguish between the gaps on the hole FSs, but cannot resolve
the two gaps on the electron FSs (a possible exception is the
hole-doped $Ba_{1-x}K_x Fe_2As_2$~\cite{122hole_ARPES2}). The
issue of the resolution is relevant because $\cos 2 \theta$
modulations of the gaps on the two electron FSs have $\pi$ phase
shift, and the convoluted (unresolved) gap remains a  constant
along the electron FS even if each of the two gaps has nodes.

Both conventional and laser ARPES measurements
\cite{ARPES_1111,122electron_ARPES,122electron_ARPES2,122hole_ARPES1,122hole_ARPES2,122hole_Matsuda,LiFeAs_ARPES1_nutron_scat,LiFeAs_ARPES2}
have shown that the gaps on the hole FSs, $\Delta_h$, are weakly
angle-dependent, but $2\Delta_h/T_c$ differs from BCS value of
$3.53$ (Ref.~\onlinecite{inosov_1}). The issue we discuss in this
communication is whether one can make a prediction, based on these
measured $2\Delta_h/T_c$, about the presence or absence of the gap
nodes on the electron FSs.  We consider 4-pocket and 5 pocket
models of FeSCs and argue that, if the largest $2\Delta_h/T_c$ is
below or above a threshold value (different for 4 and 5 pocket
models), the electron gaps either definitely have nodes, or
definitely have no nodes, respectively. There is a ``gray'' area
for $2\Delta_h/T_c$ around the threshold, when the electron gap is
either nodal or no-nodal, depending on parameters, but this gray
area is rather narrow.

{\it Method}~~~~We ignore 3D effects, potential hybridization of
the electron FSs in the folded BZ, and the difference between the
gaps due to different densities of states $N_F$ on different FSs,
and focus on the two key features associated with
multi-orbital/multi-band nature of FeSCs: the presence of multiple
FS pockets and the angle dependencies of the interactions between
low-energy fermions. The latter originates from the fact that
orbital character of low-energy states varies along the FSs. In
the band basis, this variation is passed onto the interactions
which become angle dependent. We project the interactions onto $s-$wave
channel, solve the coupled set of non-linear
gap equations for the gaps along hole and electron FSs and relate
$2\Delta_h/T_c$ to the strength of the $\cos 2 \theta$ component
of the electron gap.

We make several simplifying assumptions aiming to reduce the
number of input parameters.  First, we only keep the angular
dependence of electron-hole interaction and approximate hole-hole
and electron-electron interactions by constants. This is in line
with the earlier study~\cite{saurabh_RPA} which found that the
structure of $s^{\pm}$ gap is chiefly determined  by the angle
dependence of the electron-hole interaction $u_{eh} (\theta)$.
Second, we factorize $u_{eh}$ into an overall scale and an angular
part in which we keep the leading $2\theta$ harmonic:  $u_{eh} =
u_{eh} (1 \pm 2\alpha \cos 2 \theta)$ (for justification see
Refs.\cite{saurabh_RPA,saurabh_RG,CVV}).  Third, we take the
ratios of different intra-band and inter-band hole-hole,
electron-electron, and hole-electron interactions to be the values
to which they flow under RG~\cite{saurabh_RG}.  This leaves us
with just three parameters: the magnitude of  electron-hole
interaction, $u_{eh}$, the factor $\alpha$, which measures the
strength of its $\cos 2 \theta$ component, and the magnitude of
intra-pocket hole-hole interaction $u_{hh}$. Other hole-hole and
electron-electron interactions scale as $u_{hh}$ with the
prefactors (which are different in different models) set by the RG
flow~\cite{saurabh_RG}. The overall magnitude of the interaction
doesn't enter into $2\Delta_h/T_c$, hence the actual number of
input parameters is two: $\alpha$ and $\beta \equiv
u_{hh}/u_{eh}$. We solve the non-linear gap equations in the two
limits: when both hole FSs centered at $(0,0)$ in the unfolded BZ
are equal, and when only one FS is present.  This is done
realizing that the actual case of the two non-equivalent hole FSs
falls in between the two limits.

The gap structure for our interaction is specified by
\bea
\label{eq:gap forms}
&&\Delta^{(0,0)}_{h}(\textbf{k})=\Delta_{h_1}, ~~~\Delta^{(\pi,\pi)}_{h}(\textbf{k})=\Delta_{h_2} \nonumber\\
&&\Delta^{(\pi,0)}_{e} (\textbf{k}),~
\Delta^{(0,\pi)}_{e}(\textbf{k}) =\Delta_{e_1} \pm \Delta_{e_2}
\cos~2\theta, \eea For briefness, we discuss our computational
procedure only for the 4-pocket model with two identical hole FSs
at $(0,0)$.  The computations for other models are similar.

The set of coupled BCS-type equations at T=0 is obtained by
conventional means and reads:
 \bea
 \label{eq:4P2H Gap Eq} \Delta_h &=&
-4\beta u_{he} \Delta_h
\log\frac{2\Lambda}{|\Delta_h|} -\nonumber\\
&&2u_{he}\int\frac{d\theta}{2\pi}\left(1+2\alpha
\cos2\theta\right) \Delta_e(\theta)
\log\frac{2\Lambda}{|\Delta_e(\theta)|}\nonumber\\
\Delta_{e_1} &=& -4u_{he} \Delta_h \log\frac{2\Lambda}{|\Delta_h|}
- 4\beta u_{he}\int\frac{d\theta}{2\pi} \Delta_e(\theta)
\log\frac{2 \Lambda}{|\Delta_e(\theta)|}\nonumber\\
\Delta_{e_2} &=& -8 \alpha u_{he} \Delta_h \log\frac{2
\Lambda}{|\Delta_h|} \eea where $\Delta_e (\theta) = \Delta_{e_1}
+ \Delta_{e_2} \cos 2 \theta$ and $\Lambda$ is the upper cutoff.
We treat $u_{he}$ as dimensionless, meaning that it is the product
of the actual interaction and $N_F$.

The conventional route to find $2\Delta/T_c$ is to compare these
non-linear gap equations  with the linear ones at $T_c$, solve for
$u_{he} \log \Lambda/T_c$, substitute the result into the
non-linear gap equation and obtain the closed-form equation for
${\tilde \Delta_i} \equiv \frac{\gamma\D_i}{\pi T_c}$ ($\log
\gamma \approx 0.577$). For one-band BCS superconductor this
yields $\log{{\tilde \Delta}} =0$, i.e., $2\Delta/T_c = 2\pi
/\gamma =3.53$.  For a multi-band superconductor,  $2\Delta/T_c$
changes from 3.53 by two reasons: because hole and electron gaps
are all different, and because the ratios between the gaps change
by $O(u_{he})$ in between $T_c$ and $T=0$. Accordingly, we
introduce $\Delta_{e_1} = \gamma_1 \Delta_h, \Delta_{e_2}
=\gamma_2 \Delta_h$ and write $\gamma_1 = \gamma^{o}_1 + u_{he}
\gamma^{1}_1, \gamma_2 = \gamma^{o}_2 + u_{he} \gamma^{1}_2$,
where $\gamma^{o}_{1,2}$ are the values of $\gamma_{1,2}$ at
$T_c$. We re-express the linearized gap equations at $T_c$ as
equations on $u_{he} \log \Lambda/T_c$ and $\gamma^{o}_{1,2}$:
\beq \label{eq:4PH2_gap_matrix}\left[\textbf{1}+\left(
\begin{array}{ccc}
4\beta&2&2\alpha \\
4&4\beta&0\\
8\alpha&0&0
\end{array}\right)u_{he}L\right]
\left[
\begin{array}{c}
1\\
\gamma^o_1\\
\gamma^o_2
\end{array}\right]=0,
\eeq where $L=ln\frac{\gamma\Lambda}{\pi T_c}$. We solve (\ref{eq:4PH2_gap_matrix}), substitute the
solutions into (\ref{eq:4P2H Gap Eq}), collect terms $O(u_{he})$
 and obtain the set of three equations on ${\tilde \Delta}_h$, $\gamma^{1}_1$, and
$\gamma^{1}_2$, with $\beta$ and $\alpha$ as parameters:
 \beq
 \label{eq:4P2H GapIS_2} \left(
\begin{array}{ccc}
1&2 l&2\alpha l\\
\gamma^o_1&1+4\beta l&0\\
\gamma^o_2&0&1
\end{array}\right)
\left(
\begin{array}{c}
 a \log \tilde \D_h\\
\gamma^1_1\\
\gamma^1_2
\end{array}\right)=-\left(
\begin{array}{c}
2 \chi_1\\
4\beta \chi_2\\
0\end{array} \right) \eeq where
$\gamma^0_1$ and $\gamma^0_2$ are the solutions of  (\ref{eq:4PH2_gap_matrix}) (also functions of $\alpha$ and $\beta$ only), $a = - (8 \alpha)/\gamma^0_2$, and
\begin{eqnarray*}
\chi_1&=& \int \frac{d\theta}{2\pi}\left(1+2\alpha
\cos2\theta\right)\left(\gamma^o_1+\gamma^o_2
\cos2\theta\right) L_\gamma \nonumber \\
\chi_2&=& \int \frac{d\theta}{2\pi}\left(\gamma^o_1+\gamma^o_2
\cos2\theta\right)  L_\gamma,~L_\gamma= \log\frac{1}{|\gamma^o_1+\gamma^o_2 cos2\theta|},
\end{eqnarray*}
Solving this set we obtain ${\tilde \Delta}_h$ and  $r \equiv
|\Delta_{e_2}/\Delta_{e_1}| = \gamma_1/\gamma_2 \approx
\gamma^0_1/\gamma^0_2$
 as functions of $\alpha$ and $\beta$. Comparing the two functions, we identify the ${\tilde
\Delta}_h$ for which  $r>1$, i.e., the gap along the electron FS  has nodes.

\begin{figure}[htp]
\includegraphics[width=2.6in]{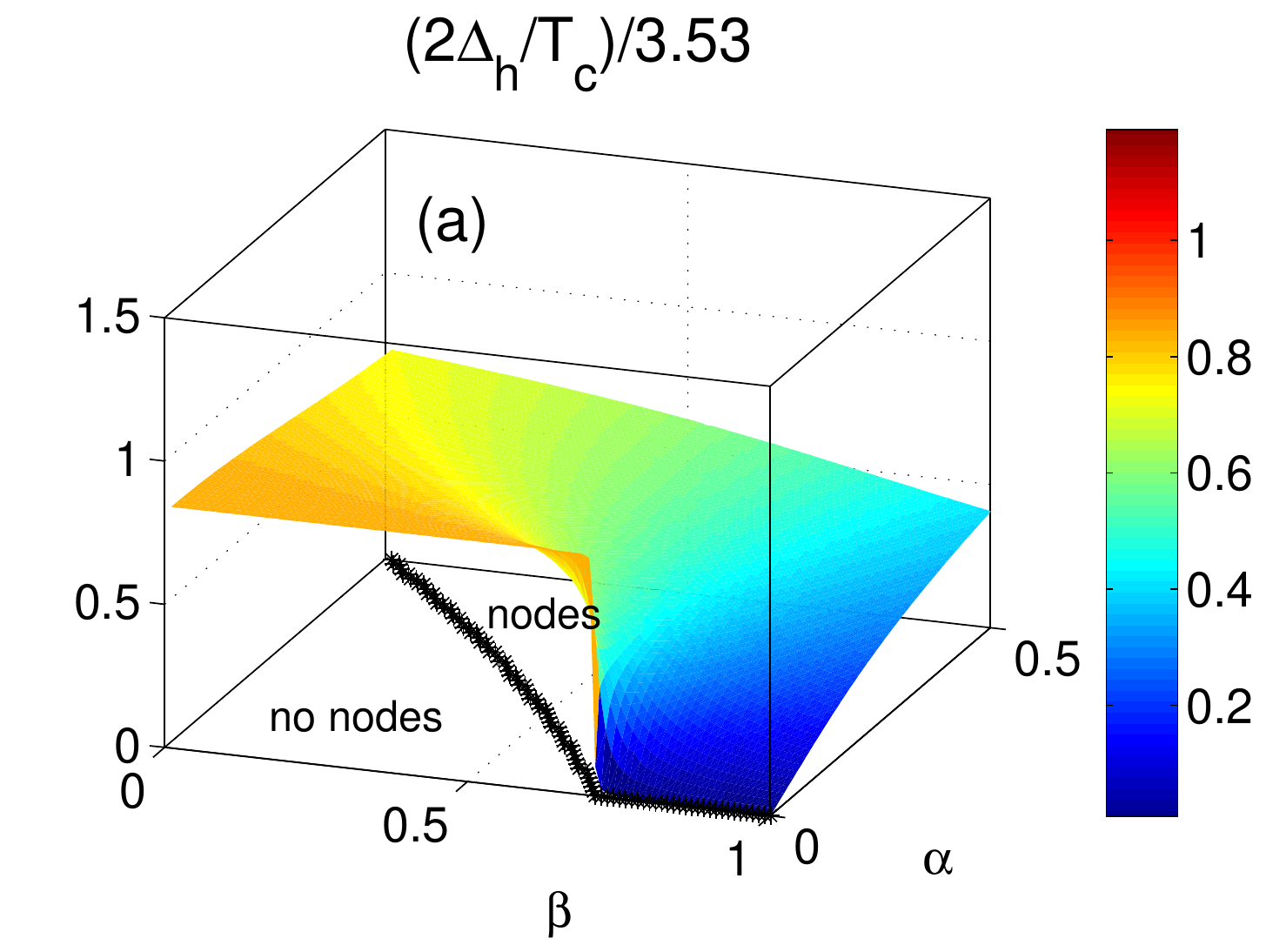}
\includegraphics[width=2.6in]{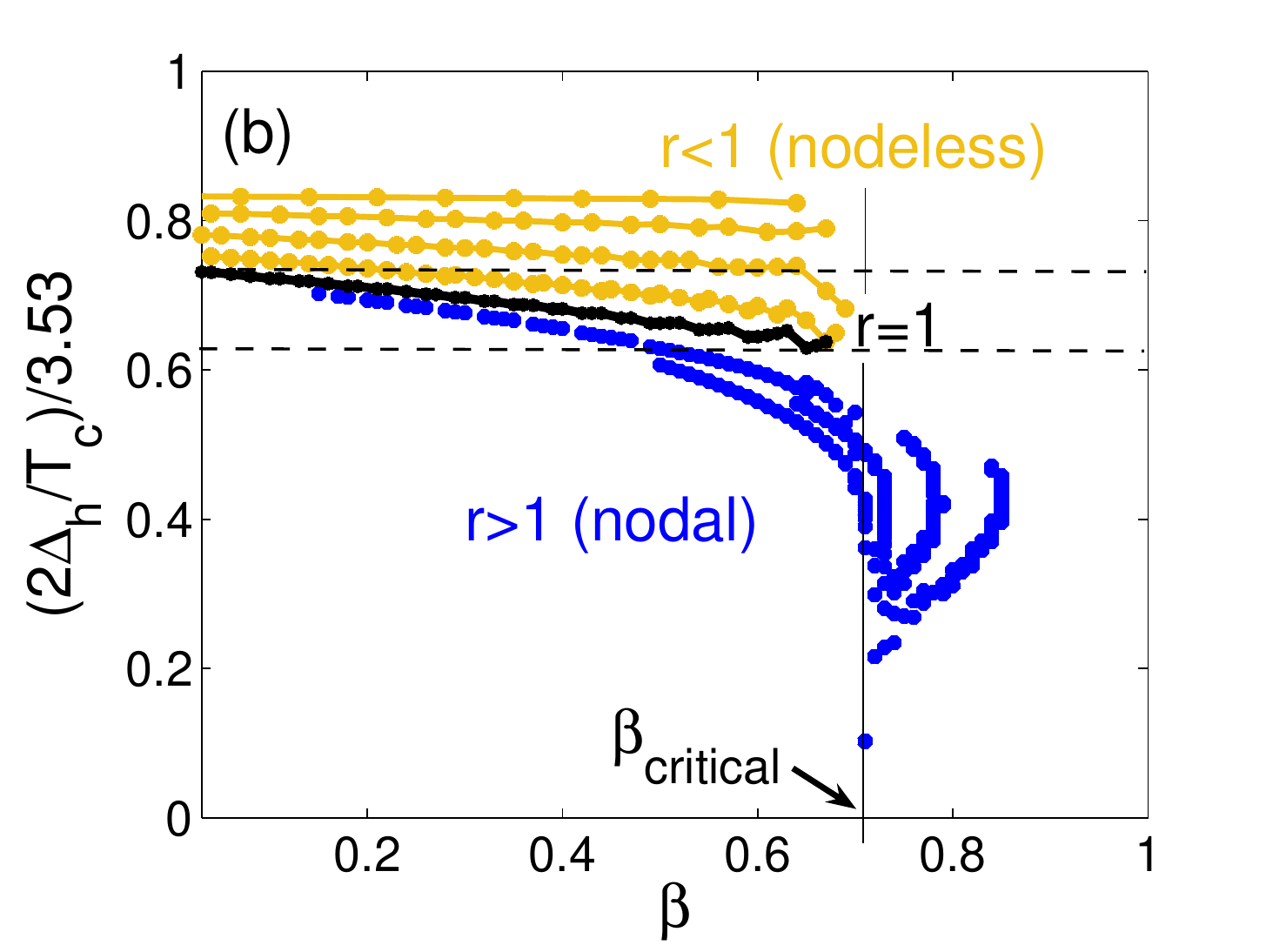}
\caption{\label{fig:deltas_4P2H} (color online) $2\D_h/T_c$ for
the gap on one of the two identical hole pockets at $(0,0)$ in the
4-pocket model. (a)  $2\D_h/T_c$ as a function of $\alpha$ and
$\beta$. The black line separates the nodal and nodeless regions.
The colors indicate the gap magnitudes relative to the BCS value
of 3.53. (b) $2\D_h/T_c$ (relative to the BCS value) vs $\beta$
for different values of $r$. $r>1$ (blue) are nodal lines, $r<1$
(yellow) are nodeless lines and $r=1$ (black) is the boundary
line. The dashed lines indicate the boundaries of the `gray area'
(see text). Outside of gray area the gap either definitely has
nodes or is definitely no-nodal.}

\end{figure}

{\it Results}~~~In Fig. \ref{fig:deltas_4P2H}a we show ${\tilde
\Delta}_h$ for different $\alpha$ and $\beta$ compared to BCS
value. The black line separates the regions of nodal and no-nodal
gap ($r >1$ and $r<1$, respectively). The nodal region corresponds
to larger values of $\alpha$ (larger angular dependence of the
interaction) and larger $\beta$(larger intra-band repulsion).
There is a critical value $\beta_{critical} = 1/\sqrt{2}$ beyond
which the gap remains nodal even when $\alpha$ is infinitesimally
small~\cite{CVV}. In Fig.\ref{fig:deltas_4P2H}b we take the slices
of Fig. \ref{fig:deltas_4P2H}a and show the trajectories of
${\tilde \Delta}_h$ for fixed values of $r$ (for every given
$\beta$, different $r$ correspond to different values of
$\alpha$). We clearly see that there is a
 correlation between the magnitude of ${\tilde \Delta}_h$ and
whether the gap along electron FS is nodal or has no nodes.
Namely, for ${\tilde \Delta}_h$ above 0.73 the gap  has no nodes,
and for ${\tilde \Delta}_h$ below 0.63, the gap has nodes, no
matter what $\alpha$ is. There is a ``gray area'' between 0.63 and
0.73 marked by dashed lines in the figure. For ${\tilde \Delta}_h$
in this area, the gap is either nodal or no-nodal, depending on
$\alpha$.

We found similar behavior for all cases that we studied. Namely,
for ${\tilde \Delta}_h$  above some threshold correspond the gap  along the electron FSs has no nodes,  for ${\tilde \Delta}_h$ below
some other threshold electronic gaps have nodes, and there is some
relatively narrow ``gray area'' in between the thresholds. In Tab.
\ref{tab:table} we summarize the main results  for  4 pocket and 5 pocket models with either two equivalent hole FSs at $(0,0)$ or only one  hole FSs at $(0,0)$ These choices are dictated by our desire to
minimize the number of input parameters and at the same time to
understand a generic case of two non-equivalent hole FSs at
$(0,0)$, which should be in between the two limits.

\begin{table}[htp]
\caption{Table summarizing the main results for (a) 4 pocket model
with 2 hole FSs at $(0,0)$, (b) 4 pocket model with 1 hole FS at
$(0,0)$, (c) 5 pocket model with 2 hole FSs at $(0,0)$, and (d) 5
pocket model with 1 hole FSs at $(0,0)$. $\Delta_h$ corresponds to
the largest hole gap in each model. The `gray area' values are
relative to the BCS value of 3.53. $2\Delta_h/T_c$ (also relative
to the BCS value) and $\Delta_{e_1}/\Delta_h$ are listed for
$\alpha\rightarrow 0, \beta\rightarrow 0$.}\label{tab:table}
\begin{ruledtabular}
\begin{tabular}{ccccc}
&$\D_{e_1}/\D_h$&$\frac{2\D_h}{T_c}/3.53$& Gray area &$\beta_{critical}$\\
\cline{2-5}
(a)&$-\sqrt{2}$&$2^{-1/4}$&$0.63-0.73$&$1/\sqrt{2}$\\
(b)&$-1/\sqrt{2}$&$2^{1/4}$&$0.94-1.04$&$\sqrt{2}$\\
(c)&$-1$&$ 2^{1/4}$&$0.84-1.04$&$0.5$\\
(d)&$-1$&$1$&$0.78-0.87 $&$1$\\
\end{tabular}
\end{ruledtabular}
\end{table}

\begin{figure*}[t]
$\begin{array}{c}
\includegraphics[width=6in]{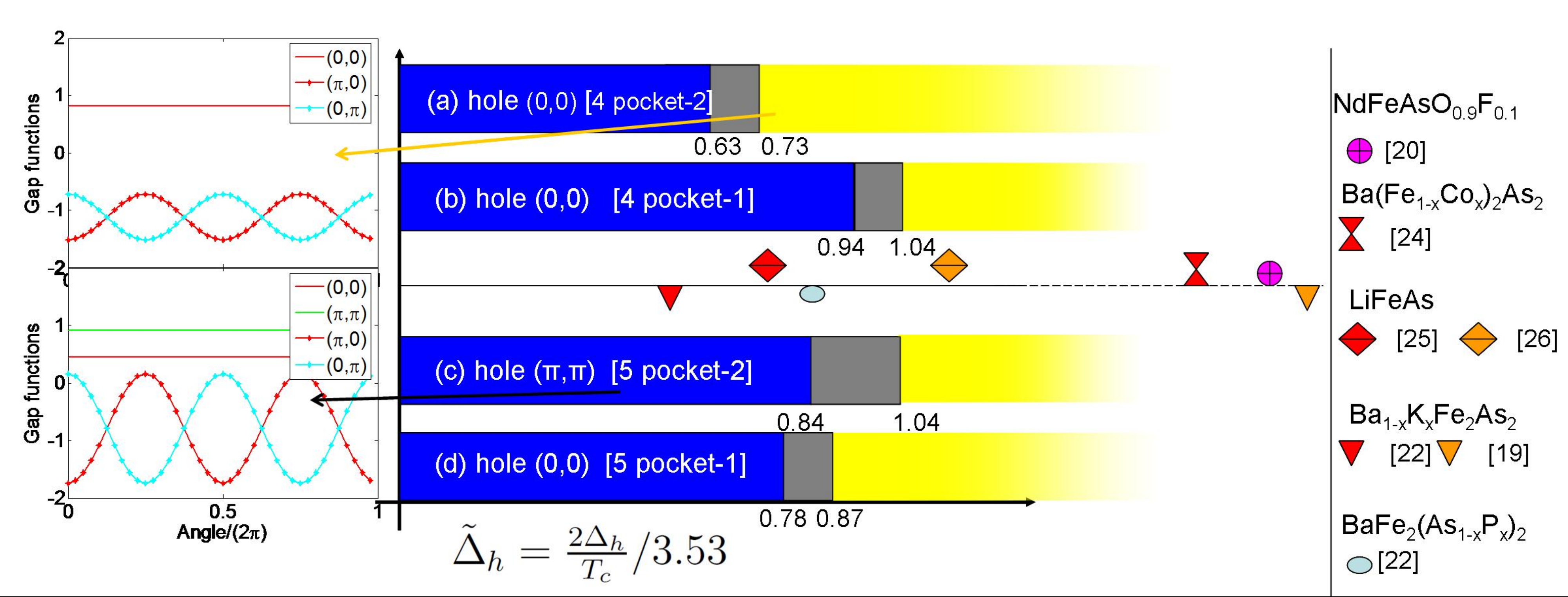}\\
\end{array}$
\caption{\label{fig:window}(color online)Gap structure strips for
(a) 4-pocket model with 2 identical hole FSs at $(0,0)$, (b)
4-pocket model with 1 hole FS at $(0,0)$, (c) 5-pocket model with
2 hole FSs at $(0,0)$ and (d) 5-pocket model with 1 hole FS at
$(0,0)$. Each strip shows the regions of $2\D_h/T_c$ for the
largest hole gap in respective models where we find nodes
(blue)/no-nodes(yellow) and the gray area (gray). The gaps are
plotted in the left panel for one nodal and one no-nodal
structure. The symbols are the ARPES values for $2\D_h/T_c$ from
various references. The symbols above the horizontal line should
be compared with the 4-pocket models and those below the line with
the 5-pocket models.}
\end{figure*}

{\it Comparison with experiments}~~~~ A summary of the results is
pictorially presented in `structure strips' in Fig.
\ref{fig:window}, where for each case we show ${\tilde \Delta}_h$
and its partition into the nodal, nodeless and the gray area. We
keep the upper boundary open because strong coupling effects are
known to increase the value of ${\tilde
\Delta}_h$~\cite{marsiglio}. The symbols represent the ARPES data
for the gaps along the hole FSs.
Since our objective is to set an upper bound on $\tilde\D_h$ for
nodal behavior, we compare the largest of the hole gaps from the
theory to the largest hole gap observed in the experiments. The
symbols above the line should be compared with the 4-pocket models
and those below the line with the 5-pocket models.

For electron-doped 122 material $Ba(Fe_{1-x}Co_x)_2As_2$, both theory and
experiment indicate that there are 2 hole and two electron
pockets. ARPES result suggests\cite{122electron_ARPES2} that one
hole pocket is near-nested with electron pockets, while other is
not. From our perspective, this should be close to our case of one
hole and two electron pockets. We see that the measured  ${\tilde
\Delta}_h$ near optimal doping sits well in the nodeless regime.
This is consistent with specific heat
and thermal conductivity data, which indicate that the gap in
optimally doped $Ba(Fe_{1-x}Co_x)_2As_2$ has no
nodes\cite{122electron_SpHeat,thermal_opt}.

A similar situation holds for electron doped 1111 material
NdFeAsO$_{0.9}$F$_{0.1}$.  The measured  $\Delta_h$ is around $15
meV$~\cite{ARPES_1111} for high $T_c \approx 53K$, which yields
$2\Delta_h/T_c \approx 6.57$, well into the nodeless region. The
no-nodal gap in high $T_c$ 1111 materials is consistent with
penetration depth measurements~\cite{pen_depth_1111} assuming that
inter-pocket impurity scattering is relevant~\cite{impurities}.

For 111 LiFeAs, which has 2 hole and 2 electron FSs, ARPES results
\cite{LiFeAs_ARPES1_nutron_scat,LiFeAs_ARPES2} suggests that the
measured  ${\tilde \Delta}_h$ is either  in the nodeless region or
near the upper boundary of gray area, suggesting that the full gap
has no nodes, if, indeed LiFeAs is an $s^{\pm}$ superconductor
(see Ref.~\onlinecite{buechner}).   The no-nodal electron gap is
consistent with penetration depth and specific heat
experiments~\cite{thermal_pen_nodeless, LiFeAs_ARPES2} which
clearly show exponential behavior at low $T$.

For hole-doped 122 material $Ba_{1-x}K_xFe_2As_2$, ARPES data show three hole FSs,
consistent with the fact that these are hole-doped materials. The
ARPES data for ${\tilde \Delta}_h$ vary. The data  by Nakayama
\emph{et. al.}~\cite{122hole_ARPES2} show that the highest
${\tilde \Delta}_h$ is rather large (about twice BCS value), which
places this material deep into the nodeless region, where the gap
along electron FSs is almost angle-independent.  There is no
evidence of nodes in this material from other
measurements.~\cite{122hole_pen_depth}

Finally, for iso-valent doping in 122 BaFe$_2$(As$_{1-x}$P$_x$)$_2$,
recent laser ARPES measurements\cite{122hole_Matsuda} detected
three near-equivalent and near-isotropic hole gaps with ${\tilde
\Delta}_h \sim 0.85$. This places the material near the lower
boundary of the gray area (see Fig.\ref{fig:window}c), i.e., from
our analysis, the measured ${\tilde \Delta}_h$ {\it implies that
there must be nodes along the electron FSs.} This is in line with
thermal conductivity and penetration depth measurements, which
show behavior consistent with the nodes in
BaFe$_2$(As$_{1-x}$P$_x$)$_2$~\cite{P_mSR_ThermalCon}. We caution,
however, that the same  laser ARPES study\cite{122hole_Matsuda}
reported  ${\tilde \Delta}_h \approx 0.52$ in
$Ba_{1-x}K_xFe_2As_2$, much smaller than other ARPES measurements.
That value would imply the nodes in  $Ba_{1-x}K_xFe_2As_2$, in
variance with what is known for this system.

{\it Conclusions}~~~~The purpose of his work was to investigate
whether one can predict,  based on the measured $2\Delta_h/T_c$
along the hole FSs, whether or not the gaps on the electron FSs in
FeSCs have nodes. The hole gaps have been measured by ARPES, but
for most systems, ARPES measurements of the electron gaps
separately on each of the two electron FSs are lacking.  This
issue is particularly relevant for systems like
BaFe$_2$(As$_{1-x}$P$_x$)$_2$, in which penetration depth and
thermal conductivity data show behavior consistent with the gap
nodes.

We considered 4-pocket and 5-pocket models with angle-dependent
interaction between hole and electron pockets and found that there
is a direct correlation between  $2\Delta_h/T_c$ on hole FSs and
the strength of $\cos 2\theta$ oscillating gap component on the
electron FSs. If  $2\Delta_h/T_c$ is larger than a certain value,
there are no nodes, if it is smaller, then there must be nodes.
There is a rather narrow range of  $2\Delta_h/T_c$ near the
boundary where the electron gap is either nodal or not depending
on the strength of the $\cos 2 \theta$ component of the
interaction.

Most of ARPES data for $\Delta_h$ are for  near-optimally doped
NdFeAsO$_{0.9}$F$_{0.1}$, $Ba(Fe_{1-x}Co_x)_2As_2$,
$Ba_{1-x}K_xFe_2As_2$, and $LiFeAs$ yield $2\Delta_h/T_c$ above
the threshold, meaning that there should be no nodes along the
electron FSs. This is consistent with the penetration depth,
thermal conductivity, and specific heat measurements in these
materials. For P-doped BaFe$_2$(As$_{1-x}$P$_x$)$_2$,
$2\Delta_h/T_c$ obtained by laser ARPES measurements is in the
lower part of the gray area,  meaning that the nodes on electron
FSs are very  likely. The detailed ARPES measurements of the
electron gaps separately on the two hole FSs are clearly called
for.

We acknowledge helpful discussions with R. Fernandes, P.
Hirschfeld, I. Eremin, Y. Matsuda, and M. Vavilov. This work was
supported by NSF-DMR-0906953.

\end{document}